\documentclass[12pt]{article}
\usepackage{amsmath}
\usepackage{amssymb}
\usepackage{epsfig}
\usepackage[latin5]{inputenc}

\numberwithin{equation}{section}
\newcommand{\dx}{\partial_x}

\newcommand{\dt}{\partial_t}

\newcommand{\dw}{\partial_\omega}
\newcommand{\dr}{\partial_{\rho}}

\newcommand{\ft}{\Phi_t}
\newcommand{\fx}{\Phi_x}
\newcommand{\gt}{\Gamma}

\newtheorem{prop}{Proposition}

\newtheorem{rmk}{Remark}

\DeclareMathOperator{\simi}{gs}

\begin{document}
\title{\bf
\Large On integrability of variable coefficient nonlinear Schr\"{o}dinger equations}

\author{
{C. \"{O}zemir} \\ \small
\small Department of Mathematics, Faculty of Science and Letters,\\
\small \.{I}stanbul Technical University, 34469 \.{I}stanbul,
Turkey
\thanks{e-mail: ozemir@itu.edu.tr}\and
{F.~G\"ung\"{o}r}\\ \small
Department of Mathematics, Faculty of Arts and Sciences,\\ \small Do\u{g}u\c{s} University, 34722 Istanbul, Turkey \thanks{e-mail: fgungor@dogus.edu.tr} }

\date{\today}

\maketitle

\begin{abstract}
We apply Painlev\'e test to the most general variable coefficient nonlinear Schr\"{o}dinger (VCNLS) equations as an attempt to identify  integrable classes
and compare our results versus those obtained by the use of other tools like group-theoretical approach and the Lax pairs technique of the soliton theory.
We present explicit transformation formulae that can be used to generate new analytic solutions of VCNLS equations  from those of the integrable NLS equation.
\end{abstract}

\section{Introduction}

Variable coefficient extensions of nonlinear evolution type
equations tend to arise in  cases when less idealized conditions
such as inhomogeneities and variable topographies are assumed in
their derivation. For example, variable coefficient Korteweg-de
Vries and Kadomtsev-Petviashvili equations describe the propagation
of waves in a fluid under the more realistic assumptions including
non-uniformness of the depth and width, the compressibility of
the fluid, the presence of vorticity and others. While these
conditions lead to variable coefficient equations, all or some of
the integrability properties of their standard  counterparts,
namely when the coefficients are set equal to constants, are in
general destroyed. However, when the coefficients are
appropriately related or have some specific form,  the generalized
equation can still be integrable as evidenced by the presence of
Painlev\'e property, Lax pairs, symmetries and other attributes of
integrability  \cite{Gungor96, Gungor01-3}. This fact suggests us
to look for a point transformation mapping VCNLS equation to the
standard NLS one.
It is thus our intention  to examine the integrability conditions  of a class of variable coefficient nonlinear Schr\"{o}dinger (VCNLS)  equation which can be thought of as a generalization of the standard integrable nonlinear  Schr\"{o}dinger (NLS) equation.
More precisely, we ask the question: Under which conditions can the equation be transformed to its constant coefficient version which is notoriously integrable?

In this paper we consider  the following
class of equations
\begin{eqnarray}\label{VCS}
\begin{split}
i\psi_t+f(x,t)\psi_{xx}+g(x,t) |\psi|^2 \psi+h(x,t) \psi=0,\\
f=f_1+{i}f_2,   \quad g=g_1+{i}g_2, \quad h=h_1+{i}h_2,\\
f_j,g_j,h_j\in\mathbb{R}, \quad j=1,2, \quad f_1\neq0, \quad
g_1\neq0
\end{split}
\end{eqnarray}
and study its integrability and in particular reducibility to its standard form using the tools of singularity analysis and symmetry which are well-established approaches towards integrability.
This class of equations model a large variety of physical phenomena and has been widely
analyzed through various methods. The complex function $\psi(x,t)$ has different physical meanings in different physical contexts. An equation with an additional quintic nonlinearity when the coefficient
functions mostly depend on time models the propagation of  pulses in optic fibers
 and was studied in \cite{Zhu09}, where  trigonometric type solutions were obtained through some transformations.
  \cite{Hao04} studies the equation with space and time coordinates switched, modeling propagation of pulses
  in optical fibers with distributed dispersion and nonlinearity and finds soliton-type solutions with a Darboux
  transformation, whereas elliptic-type solutions through various transformations \cite{Zhang07}, soliton-type
  solutions via Hirota method \cite{Lu07} and in terms of a double-Wronskian determinant \cite{Lu08},
  and self-similar solutions also exist in the literature \cite{Kruglov03}. In these works,
  the coefficients are considered as functions of a single variable. For another subclass of equations when $f,g$ are real
  functions of $t$, reductions to a nonlinear ODE by an appropriate transformation is performed and
  solutions of hyperbolic, trigonometric and elliptic-type are obtained \cite{Liu07}. Besides,
  elliptic-type solutions by a direct-symmetry method are found in \cite{Yan07} for the same case.
  \cite{Liu04} considers the case with constant $f,g$ and a real potential periodic in time,
  and presents soliton-type solutions. For $f=1$ and $g,h$ real, \cite{Beitia08} uses similarity
  transformations to transform the equation into the stationary nonlinear Schr\"{o}dinger equation
   and hence obtains soliton-type solutions of various characters.

 For  $f,g$ real function of $t$ and a real potential which is quadratic in $x$ with
 time-dependent coefficients, \cite{Serkin07} gives  an integrability condition
 based on a Lax pair. \cite{Gurses08} tries to transform this equation directly
 to the NLS equation, whereas \cite{Kundu09} states that this equation and its extended
 version are equivalent to the NLS equation.

 There is a vast amount of literature devoted to analyzing Painlev\'e property (P-property)
 of equations belonging to the class \eqref{VCS} as subcases (for example, see \cite{Steeb84, ZLC08}).
 Results of \cite{ZHL08} and \cite{LZHL08} belong to a wider
 class and the compatibility condition they find coincides with
 the exact integrability condition given by \cite{Serkin07}.
 Finally, we would like to mention Ref. \cite{Winternitz93},
 in which allowed transformations of \eqref{VCS}
  are found and used to do a complete symmetry group
   analysis of the equation.

In addition, we apply the Painlevé test to  the  more general equations
\begin{equation}\label{VCSG}
i\psi_t+f\,\psi_{xx}+g\, |\psi|^2 \psi+h\, \psi+k\,\psi_x=0\\
\end{equation}
and
\begin{equation}\label{VCSG2}
i\psi_t+f\,\psi_{xx}+g\, |\psi|^2 \psi+h\, \psi+k\,\psi_x+i\, l\,\psi_{xxx}+i\, m\,(|\psi|^2 \psi)_x+i\, n \, \psi (|\psi|^2)_x=0,\\
\end{equation}
where the coefficient functions are considered to be arbitrary in the form  $f=f_1(x,t)+i\,f_2(x,t)$, $g=g_1(x,t)+i\,g_2(x,t)$, $h=h_1(x,t)+i\,h_2(x,t)$, $k=k_1(x,t)+i\,k_2(x,t)$,
$l=l(t)$, $m=m(t)$, $n=n(t)$. P-test results for \eqref{VCSG} are obtained in   \cite{Brugarino10} when $f$ is real. In Ref. \cite{Li07}  integrability of
\eqref{VCSG2} is investigated for the coefficients having only time dependence. Here we allow also spatial dependence and obtain the Painlevé-integrability conditions.
We refer the interested reader to these works  for the physical motivation of these generalized equations.

 The paper is organized as follows.
 Section 2 performs the Painlev\'e analysis of
 equations \eqref{VCS}-\eqref{VCSG2}. Section 3 is devoted to the transformation of
  the equation into the NLS equation using allowed transformations.
We end up by applying our results to a generalized Gross-Pitaevskii equation in a harmonic trap
   in Section 4.

\section{Singularity Analysis of the Equations}

\subsection{Painlevé test for eq. \eqref{VCS} }
For convenience we write \eqref{VCS} together with its complex conjugate as the
system
\begin{eqnarray}\label{sys}
\begin{split}
iu_t+f(x,t)u_{xx}+g(x,t) u^2 v+h(x,t) u=0,\\
-iv_t+p(x,t)v_{xx}+q(x,t) u v^2+r(x,t) v=0.\\
\end{split}
\end{eqnarray}
Here $u$ was employed instead of $\psi$ and $v$ denotes its
complex conjugate, but in this context they are viewed as
independent functions, whereas $p,q,r$ are complex
conjugates of $f,g$ and $h$, respectively. This system will be subjected to the Painlev\'e test for partial differential equations. We shall look for
solutions of the system ~\eqref{sys} in the form of a Laurent expansion (known as Painlev\'e expansion) in the neighborhood of a non-characteristic, movable
singularity surface (actually a  curve in this case) defined by
\begin{equation}\label{3.2}
\Phi(x,t)=0.
\end{equation}
Thus, we expand
\begin{equation}\label{exp}
u(x,t)=\sum_{j=0}^{\infty} \; u_j(x,t)\Phi^{\alpha+j}(x,t), \quad
v(x,t)=\sum_{j=0}^{\infty} \; v_j(x,t)\Phi^{\beta+j}(x,t),
\end{equation}
where  $u_0, v_0\ne 0$ and $u_j, v_j, \Phi(x,t)$ are analytic
functions. $\alpha$ and $\beta$ are negative integers to be determined from
the leading order analysis.

The partial differential equation (PDE) is said to pass the
Painlev\'e test if the substitution of the above expansion into the
equation leads to the correct number of arbitrary functions as
required by the Cauchy-Kovalewski theorem given by the expansion
coefficients, where $\Phi(x,t)$ should be one of the arbitrary
functions. The coefficients in the expansion where the arbitrary
functions figure are known as the resonances.

If a PDE passes this test, then it has a good chance of having the
Painlev{\'e} property: solutions \eqref{exp} are single valued
about the singularity surface, and depends on sufficient number of
arbitrary functions which are needed to satisfy arbitrary Cauchy
data imposed for some $x=x_0$. We note that passing the
Painlev{\'e} test is necessary, but not sufficient for having the
Painlev{\'e} property. In that case, for integrable PDEs it is
usually possible  to construct auto-B\"acklund transformations
which relate equations to themselves via differential
 substitutions and also Lax pairs, which then ensures the sufficient condition of integrability.
 Application of the Painlev\'e expansion to nonintegrable PDEs can allow particular explicit solutions
 to be obtained by truncating the expansion which then imposes constraints on the arbitrary functions
 and the function $\Phi$. This usually requires compatibility of an overdetermined
PDE system.  Let us mention that the method of
 truncated expansion has been successfully applied to many nonintegrable PDEs in  constructing  exact solutions.

For the determination of leading orders $\alpha$ and $\beta$, we
substitute $u\sim u_0 \Phi^\alpha$ and $v\sim v_0\Phi^\beta$ in
\eqref{sys} and see that by balancing the terms of
smallest order
\begin{equation}
\alpha+\beta=-2
\end{equation}
must hold, which only allows the negative integers $\alpha=-1$ and
$\beta=-1$. In addition, we obtain the relations
\begin{subequations}\label{ini}
\begin{equation}
u_0(gu_0v_0+2f\Phi_x^2)=0
\end{equation}
\begin{equation}
v_0(qu_0v_0+2p\Phi_x^2)=0
\end{equation}
\end{subequations}
and it follows that
\begin{equation}\label{u0v0}
u_0v_0=-2\Phi_x^2\frac{f}{g}=-2\Phi_x^2\frac{p}{q}.
\end{equation}
From this it is seen that $\frac{f}{g}$ must be real, which
requires that $f_1g_2=f_2g_1$. We can say that, if one of $f$ and
$g$ is real or pure imaginary, then so is the other.

After determination of the leading orders, we substitute
\eqref{exp} into \eqref{sys}. For $j=0$, equating the coefficient
of $\Phi^{-3}$ to zero we exactly obtain the relations
\eqref{ini}. For $j\geq1$, equating coefficient of $\Phi^{-3+j}$
to zero yields the system
\begin{equation}\label{det}
Q(j)\left( \begin{array}{c}
        u_j\\
        v_j
       \end{array}\right)=
\left( \begin{array}{cc}
        (j^2-3j-2)f\Phi_x^2&gu_0^2\\
        qv_0^2&(j^2-3j-2)p\Phi_x^2
       \end{array}
\right) \left( \begin{array}{c}
        u_j\\
        v_j
       \end{array}
\right)= \left( \begin{array}{c}
        F_j\\
        G_j
       \end{array}\right),
\end{equation}
where
\begin{eqnarray}\label{sis}
\begin{split}
F_j=&-iu_{j-2,t}-i(-2+j)\phi_t u_{j-1}-fu_{j-2,xx}- 2(-2+j)f\Phi_x
u_{j-1,x}\\
&-(-2+j)f\Phi_{xx}u_{j-1}-hu_{j-2}
-gv_0\sum_{l=1}^{j-1}u_lu_{j-l}-g\sum_{k=1}^{j-1}\sum_{l=0}^{k}u_lu_{k-l}v_{j-k},\\
 G_j=\,&iv_{j-2,t}+i(-2+j)\phi_t v_{j-1}-pv_{j-2,xx}- 2(-2+j)p\Phi_x
v_{j-1,x}\\
&-(-2+j)p\Phi_{xx}v_{j-1}-rv_{j-2}
-qu_0\sum_{l=1}^{j-1}v_lv_{j-l}-q\sum_{k=1}^{j-1}\sum_{l=0}^{k}v_lv_{k-l}u_{j-k}.
\end{split}
\end{eqnarray}
Here we have  used the convention that $u_j=0$ for $j<0$ and sums
which have upper indices that are less than the lower indices will
be taken as zero. Notice that $F_j$ and $G_j$ can be obtained from
each other by the interchange $i\leftrightarrow -i$,
$u\leftrightarrow v$, $f\leftrightarrow p$, $g\leftrightarrow q$,
$h\leftrightarrow r$. $u_j$ and $v_j$  are determined by the
system \eqref{det} unless
\begin{equation}
\det Q(j)=|f|^2\Phi_x^4(j+1)j(j-3)(j-4)=0.
\end{equation}
Hence we find the resonance levels to be $j=-1,0,3,4.$ The
resonance $j=-1$ corresponds to the arbitrariness of the
singularity manifold $\Phi$ and the resonance $j=0$ points out
that there is one arbitrary function among $u_0$ and $v_0$, as it
is verified by \eqref{u0v0}.

For $j=1$ we solve
\begin{equation}
\left( \begin{array}{cc}
        -4f\Phi_x^2&gu_0^2\\
        qv_0^2&-4p\Phi_x^2
       \end{array}
\right) \left( \begin{array}{c}
        u_1\\
        v_1
       \end{array}
\right)= \left( \begin{array}{c}
        i\ft u_0+2f \fx u_{0,x}+f\Phi_{xx}u_0\\
        -i\ft v_0+2p\fx v_{0,x}+p\Phi_{xx} v_0
       \end{array}\right)\end{equation}
and find the expansion coefficients to be
\begin{eqnarray}
\begin{split}
u_1=&-\frac{2f\fx
v_{0,x}}{gv_0^2}+\frac{4f_x\fx}{3gv_0}-\frac{4fg_x\fx}{3g^2v_0}+
\frac{2i\Phi_t}{3gv_0}+\frac{i\Phi_t}{3qv_0}+\frac{3f\Phi_{xx}}{gv_0},\\
v_1=&-\frac{v_{0,x}}{\fx}+\frac{f_xv_0}{3f\fx}-\frac{g_xv_0}{3g\fx}+\frac{i\Phi_tv_0}{6f\fx^2}
+\frac{ig\Phi_tv_0}{3fq\fx^2}+\frac{\Phi_{xx}v_0}{2\fx^2};
\end{split}
\end{eqnarray}
making use of the facts that $u_0=-2\Phi_x^2\frac{f}{gv_0}$ and
$p=q\frac{f}{g}$. For $j=2$ we solve
\begin{equation}
\left( \begin{array}{cc}
        -4f\fx^2&gu_0^2\\
        qv_0^2&-4p\fx^2
       \end{array}
\right) \left( \begin{array}{c}
        u_2\\
        v_2
       \end{array}
\right)= \left( \begin{array}{c}
        -hu_0-gu_1^2v_0-2gu_0u_1v_1-iu_{0,t}-fu_{0,xx}\\
        -rv_0-qv_1^2u_0-2qv_0v_1u_1+iv_{0,t}-pv_{0,xx}
       \end{array}\right)\end{equation}
and find the expansion coefficients $u_2,v_2$. However, we do not
reproduce them here as they are so lengthy.

At the resonance level $j=3$, we have a linear relation between
$u_3,v_3$ and one of them is arbitrary. We see that the system
\begin{equation}\label{sys3}
\left( \begin{array}{cc}
        -2f\fx^2&gu_0^2\\
        qv_0^2&-2p\fx^2
       \end{array}
\right) \left( \begin{array}{c}
        u_3\\
        v_3
       \end{array}
\right)= \left( \begin{array}{c}
        F_3\\
        G_3
       \end{array}\right)\end{equation}
gives the compatibility condition at $j=3$ as
\begin{equation}\label{AA}
qv_0F_3=gu_0G_3.
\end{equation}
Since the compatibility condition has to be satisfied for any
choice of the manifold $\Phi$, coefficients of
$\ft,\ft^2,\ft^3,\Phi_{tt}$ occurring in \eqref{AA} must vanish
identically. From the coefficient of $\Phi_{tt}$ we find that
$g^2=q^2$. It follows that $g=q$ or $g=-q$. The first is possible
when $g=g_1$ and for the second we have $g=ig_2$. Since $f$ and
$g$ are both real or pure imaginary, we have $f=f_1$ for the
former and $f=if_2$ for the latter.

We consider two cases:

\noindent (i) The case  $g=ig_2$, $f=if_2$. From the coefficient of $\ft^2$
we find that
\begin{equation}
2\frac{v_{0,x}}{v_0}+\frac{g_{2,x}}{g_2}-\frac{f_{2,x}}{f_2}-2\frac{\Phi_{xx}}{\fx}=0,
\end{equation}
which is a restriction on $v_0$ that contradicts its
arbitrariness. We conclude that when $f$ and $g$ are pure
imaginary, the equation cannot pass the Painlev\'e test.

\noindent (ii) The case  $g=g_1$, $f=f_1$. From now on, we are going to drop
the subscript $1$ and assume that $f$ and $g$ are real functions.
Equation \eqref{AA} is satisfied if and only if
\begin{subequations}\label{AA1}
\begin{equation}\label{AA11}
\left(\frac{f_{x}}{f}\right)^2+\frac{f_{x}g_{x}}{fg}+4\left(\frac{g_{x}}{g}\right)^2-\frac{f_{xx}}{f}
-2\frac{g_{xx}}{g}=0,\\
\end{equation}
\begin{equation}\label{AA12}
h_2\frac{f_{x}}{f}+2h_2\frac{g_{x}}{g}-3h_{2,x}+\frac{3}{2}\frac{f_{t}f_{x}}{f^2}
-\frac{1}{2}\frac{f_{x}g_{t}}{fg}
+\frac{f_{t}g_{x}}{fg}-2\frac{g_{t}g_{x}}{g^2}-\frac{f_{xt}}{f}+\frac{g_{xt}}{g}=0.
\end{equation}
\end{subequations}

Next we proceed to the resonance level $j=4$. From the system
\begin{equation}
\left( \begin{array}{cc}
        2f\fx^2&gu_0^2\\
        qv_0^2&2p\fx^2
       \end{array}
\right) \left( \begin{array}{c}
        u_4\\
        v_4
       \end{array}
\right)= \left( \begin{array}{c}
        F_4\\
        G_4
       \end{array}\right)\end{equation}
we see that one of $u_4$ and $v_4$ should be arbitrary  and it
yields the compatibility condition
\begin{equation}\label{BB}
v_0F_4+u_0G_4=0.
\end{equation}
This relation contains the functions $u_3, \, v_3$ of the
resonance $j=3$. From \eqref{sys3} we substitute $u_3=(g u_0^2
v_3-F_3)/(2f\fx^2)$ in \eqref{BB}. Similar to the previous
resonance, from the coefficient of $\Phi_{tt}$ we get
\begin{equation}\label{fxgx}
\frac{f_{x}}{f}+2\frac{g_{x}}{g}=0,
\end{equation}
which integrates to give
\begin{equation}\label{fg2}
f(x,t)g^2(x,t)=K(t),
\end{equation}
where $K(t)$ is an arbitrary integration constant. Notice that
this condition is already satisfied  for the special case when $f$
and $g$ are only functions of $t$. We set
$f(x,t)={K(t)}/{g^2(x,t)}$ and under this constraint see that
\eqref{AA11} is satisfied whereas \eqref{AA12} simplifies as
$h_{2,x}=({g_t}/{g})_x$ and determines the complex
part of the potential
\begin{equation}\label{h2}
h_2(x,t)=\frac{g_t}{g}+\gamma(t),
\end{equation}
where $\gamma(t)$ is an arbitrary function. Notice that $h_2$ may
depend on $x$ if $g$ depends on $x$. The only remaining equation
that has to be satisfied is \eqref{BB}:
\begin{eqnarray}\label{h1xx}
\begin{split}
\frac{2K}{g}\left(\frac{h_{1,xx}}{g}-\frac{g_xh_{1,x}}{g^2}\right)
&=\;4\gamma^2+2\dot{\gamma}+2\gamma\frac{\dot{K}}{K}+\frac{\ddot{K}}{K}-\left(\frac{\dot{K}}{K}\right)^2
+\frac{\dot{K}}{K}\frac{g_t}{g}-\frac{g_{tt}}{g}\\
&+K^2\left(-36\frac{g_x^4}{g^8}
+48\frac{g_x^2g_{xx}}{g^7}-6\frac{g_{xx}^2}{g^6}-10\frac{g_x
g_{xxx}}{g^6}+\frac{g_{xxxx}}{g^5}\right).
\end{split}
\end{eqnarray}
It is possible to integrate this expression and find real part of the potential as
\begin{equation}\label{h1}
h_1(x,t)=\;\frac{\tau_0}{2}\left(\int g dx\right)^2+\int g(\int\tau_1 dx )dx+\frac{K}{2}\left(\frac{g_{xx}}{g^3}-\frac{3}{2}\frac{g_x^2}{g^4}\right)
+\delta_0 \int g dx+\delta_1,
\end{equation}
where $\tau_0$ and $\tau_1$ denote
\begin{equation}
\tau_0=\tau_0(t)=\frac{2\gamma^2}{K}+\frac{(\gamma
K)_t}{K^2}+\frac{(\ln K)_{tt}}{2K}, \qquad
\tau_1=\tau_1(x,t)=-\frac{1}{2}\left(\frac{g_t}{K}\right)_t.
\end{equation}
Here $\delta_0(t)$ and $\delta_1(t)$ are arbitrary functions. Note
that integrals in \eqref{h1} will be evaluated without additional
integration constants.

\begin{prop} Equation \eqref{VCS} passes the Painlev\'e test if and only if the conditions
\eqref{fg2}, \eqref{h2} and \eqref{h1} hold.
\end{prop}
We now consider a somewhat simpler case. We see from \eqref{fxgx} that if
one of $f$ and $g$ is only a function of $t$, then so is the
other. Furthermore, we do not have a condition like \eqref{fg2}
for $f$ and $g$, they can
be independent. For $f=f(t),\;g=g(t)$, \eqref{AA11} holds
identically and from \eqref{AA12} we have $h_{2,x}=0$, which means
\begin{equation}
h_2(x,t)=\beta(t).
\end{equation}
Similarly, we are left with a single equation from\eqref{BB}:
\begin{equation}\label{ref}
2fh_{1,xx}=4\beta^2+2\dot{\beta}+2\beta\frac{\dot{f}}{f}-4\beta
\frac{\dot{g}}{g}-(\frac{\dot{f}}{f})^2-\frac{\dot{f}}{f}\frac{\dot{g}}{g}+2(\frac{\dot{g}}{g})^2
+\frac{\ddot{f}}{f}-\frac{\ddot{g}}{g}.
\end{equation}
This relation also follows directly from \eqref{h1xx} with the
substitution $\gamma(t)=\beta(t)-\frac{\dot{g}}{g}$. We can write
the right hand side as
\begin{equation}
2fh_{1,xx}=4\beta^2+2\dot{\beta}+2\beta\frac{\dot{f}}{f}-4\beta\frac{\dot{g}}{g}+(\ln
f)_{tt}+g(\frac{1}{g})_{tt}-(\ln f)_{t}(\ln
g)_{t},
\end{equation}
where the subscript $t$ denotes the derivative, and integrate to obtain $h_1(x,t)$:
\begin{equation}\label{pot}
h_{1}(x,t)=\frac{1}{4f}\{4\beta^2+2\dot{\beta}+2\beta\frac{\dot{f}}{f}-4\beta\frac{\dot{g}}{g}+(\ln
f)_{tt}+g(\frac{1}{g})_{tt}-(\ln f)_{t}(\ln
g)_{t}\}x^2+H_1(t)x+H_2(t),
\end{equation}
with $H_1(t)$ and $H_2(t)$ arbitrary. This conclusion thoroughly coincides  with those of \cite{ZHL08, Zhao09}. For $f=1$ and  $g=-2$ we have
\begin{equation}
h_{1}(x,t)=\frac{1}{2}(2\beta^2+\dot{\beta})x^2+H_1(t)x+H_2(t)
\end{equation}
and it is compatible with the results of \cite{Clarkson93}.

\subsection{Painlevé test for eq. \eqref{VCSG} }

We emphasize that there is no extra difficulty in applying Painlev\'e test
 to other  variants of \eqref{VCS}. One of the physically important possible extensions is \eqref{VCSG},
where  $f,g,h,k$ are as before complex-valued functions. For the special cases of the coefficients
the equation is referred to as NLS equation with effective mass.
If  the singularity analysis is repeated for \eqref{VCSG}, it can be shown that passing the
P-test is only possible when functions $f,g$ are real  and  the following conditions hold:
\begin{align}\label{v1}
k_1(x,t)=&\frac{1}{2}f_x(x,t)+f(x,t)\frac{g_x(x,t)}{g(x,t)}\\
h_2(x,t)=&\frac{1}{4}\left((\ln \frac{g^2}{f})_t+k_2 (\ln \frac{g^2}{f})_x+2k_{2,x}\right)+\gamma(t),
\end{align}
and
\begin{eqnarray}\label{v3}
\begin{split}
&f_x h_{1,x}+2 f h_{1,xx}=4 \gamma^2+2 \dot{\gamma}-\frac{3f_t^2}{4f^2}+\frac{f_{tt}}{2f}-\frac{3 k_2 f_t f_x}{2f^2}+\frac{k_{2,t}f_x}{2f}-\frac{3 k_2^2 f_x^2}{4 f^2}\\
                         &+\frac{k_{2,x} f_t }{f} +\frac{3_2 k_{2,x} f_x}{2f}-k_{2,x}^2+\frac{k_2 f_{xt}}{f}-k_{2,xt}
                         +\frac{k_2^2 f_{xx}}{2 f}-k_2 k_{2,xx}\\
                         &-\frac{f_x^2 g_x^2}{2g^2}+\frac{7 f f_x g_x^3}{2 g^3}-\frac{3 f^2 g_x^4}{g^4}+
                         \frac{f_x g_x f_{xx}}{4g}-\frac{3 f f_{xx} g_x^2}{2 g^2}+\frac{3 f_x^2 g_{xx}}{4g}-\frac{13 f f_x g_x g_{xx}}{2 g^2}\\
                         &+\frac{7 f^2 g_x^2 g_{xx}}{g^3}
                         +\frac{2 f f_{xx} g_{xx}}{g}-\frac{2 f^2 g_{xx}^2}{g^2}+\frac{f g_x f_{xxx}}{2 g}+\frac{3 f f_x  g_{xxx}}{g}-\frac{3 f^2 g_x g_{xxx}}{g^2}+\frac{f^2 g_{xxxx}}{g}.
\end{split}
\end{eqnarray}
Here,  $k_1(x,t)$ and $k_2(x,t)$ are real and imaginary parts of $k(x,t)$  and $\gamma(t)$ is arbitrary.
Two successive integrations of
\eqref{v3} lead to the real part of the potential:
\begin{eqnarray}\label{H1}
\begin{split}
h_1(x,t)=&\int \frac{1}{\sqrt{|f|}} \, \int F_{\varepsilon}(x,t) dx \, dx+\frac{f_x g_x}{4g}-\frac{f g_x^2}{4 g^2}+\frac{f g_{xx}}{2g}\\
         &+\int \frac{k_2}{2f}\left((\ln \frac{|f|}{k_2})_t+k_2 (\ln \frac{\sqrt{|f|}}{k_2})_x\right) \, dx+A(t)\int \frac{dx}{\sqrt{|f|}}+B(t).
\end{split}
\end{eqnarray}
Here $A(t)$ and $B(t)$ are arbitrary functions and
\begin{equation}\label{FE}
F_\varepsilon(x,t)=\frac{\varepsilon}{\sqrt{|f|}}\left(2\gamma^2+\dot{\gamma}-\frac{3f_t^2}{8f^2}+\frac{f_{tt}}{4f}\right)
\end{equation}
where $\varepsilon=1$ for  $f>0$ and $\varepsilon=-1$ for  $f<0$.

We finally remark that at the time this work has been finalized it came to our attention that the recent paper  \cite{ Brugarino10} treated the variant of this equation in which $f$ is real.

\subsection{Painlevé test for eq. \eqref{VCSG2} }

We assume that the coefficients $l(t), m(t), n(t)$ are not identically zero. Furthermore, leading order analysis shows that $3m(t)+2n(t)$ cannot vanish identically, since otherwise we would have $l(t)\equiv0$. We obtain integer resonances from the Painlevé series expansion if (i) $n(t)=-m(t)$ or (ii) $n(t)=-\frac{1}{2}m(t)$.

(i) We find the resonances as  $j=\{-1,0,1,3,4,5\}$. Compatibility conditions at   the resonances hold if

\begin{align}
&f_1(x,t)=\frac{3l}{m}\, g_1, \quad f_2(x,t)=0,  \quad g_2(x,t)=0\\
&k_1(x,t)=\frac{3l}{m}\,g_{1,x}\,, \quad k_2(x,t)=-\frac{3l}{m^2}\,g_1^2+x\gamma_1+\gamma_2\\
\label{hhh}
&h_1(x,t)=\int\Big(\frac{g_{1,t}}{m}-\frac{\dot{m}g_1}{m^2}\Big)dx+\big(x\gamma_1+\gamma_2\big)\frac{g_1}{m}-\frac{l}{m^3}\,g_1^3+\frac{l}{m}\,g_{1,xx}
+x\beta_1+\beta_2\\
&h_2(x,t)=\frac{\dot{m}}{2m}-\frac{\dot{l}}{2l}-\frac{3l}{m^2}\,g_1\,g_{1,x}+\gamma_1
\end{align}
where $\gamma_i, \beta_i, i=1,2$ are arbitrary functions of time. There is no special condition on $g_1(x,t), m(t)$ and $l(t)$.

(ii) Resonances occur at $j=\{-1,0,2,3,4,4\}$. We obtain the same conditions of (i) with $\beta_1(t)=0$ in \eqref{hhh}.

We stress that these results generalize those found in \cite{Li07}
in a consistent manner. For the case (i), considering the above
results with $\gamma_1(t)=\beta_1(t)=0$ and neglecting the
indefinite integral in \eqref{hhh}, one obtains exactly the same
conditions as in the presence of time dependent coefficients. The
same arguments  lead to  the conditions for (ii) in the
time-dependent case, with the additional equation $g_1=s_0 m$,
where $s_0\neq0$  is an arbitrary constant.

\section{Transformation to the Standard NLS Equation by Allowed Transformations}
In \cite{Winternitz93} the authors give a classification of the
symmetry algebras of \eqref{VCS} into canonical forms using the
allowed (or equivalence) transformations of the form
\begin{equation}
\psi(x,t)=Q(x,t)\tilde{\psi}(\tilde{x},\tilde{t}),  \qquad \tilde{x}=X(x,t), \qquad \tilde{t}=T(t),
\end{equation}
which leaves form of \eqref{VCS} invariant, that is, the
transformation does not add any term to the equation but may
change the coefficient functions. Here $Q(x,t)$ is a complex
function obeying the constraint
\begin{equation}\label{qdenk}
i QX_t+f(x,t)(X_{xx}Q+2Q_xX_x)=0.
\end{equation}
If we set $Q(x,t)=R(x,t)e^{i \theta(x,t)}$, $R\geq0,
0\leq\theta<2\pi$, the coefficient functions $f,g$ and $h$ map
into
\begin{equation}\label{fg}
\tilde{f}(\tilde{x},\tilde{t})=f(x,t)\frac{X_x^2}{\dot{T}}, \qquad
\tilde{g}(\tilde{x},\tilde{t})=g(x,t)\frac{R^2(x,t)}{\dot{T}(t)},
\end{equation}
\begin{eqnarray}\label{hcizgi}
\begin{split}\label{h}
\tilde{h}(\tilde{x},\tilde{t})&=\frac{1}{\dot{T}}\{h_1(x,t)-\theta_t+f(x,t)\left(\frac{R_{xx}}{R}-\theta_x^2\right) \\
&+i\left[h_2(x,t)+\frac{R_t}{R}+f(x,t)(\frac{2R_x\theta_x}{R}+\theta_{xx})\right]
\}
\end{split}
\end{eqnarray}
through this transformation. Note that here $f(x,t)$ is a
complex-valued function. However, when writing $\tilde{h}$ we have
not separated it into real and imaginary parts, since from now on it is going to be taken as a real function.

One of the main results of \cite{Winternitz93} is that any
equation of the form \eqref{VCS} with a five-dimensional symmetry
algebra is equivalent to the NLS equation with $\tilde{f}=1$,
$\tilde{g}=\epsilon+i\tilde{g_2}$, $\tilde{h}=0$, where
$\epsilon=\pm1$ and $\tilde{g_2}=\mathrm{constant}$. The algebra
is solvable and has a basis
\begin{equation}\label{alg}
P_0=\partial_{\tilde{t}},  \quad P_1=\partial_{\tilde{x}}, \quad
W=\partial_{\tilde{\omega}}, \quad
B=\tilde{t}\partial_{\tilde{x}}+\frac{1}{2}\tilde{x}\partial_{\tilde{\omega}},
 \quad D=\tilde{t}\dt+\frac{1}{2}\tilde{x}\partial_{\tilde{x}}-\frac{1}{2}\tilde{\rho}\partial_{\tilde{\rho}},
\end{equation}
which is isomorphic to the one-dimensional extended Galilei
similitude algebra $\simi(1)$. Here $\psi$ is expressed in terms
of the modulus and the phase of the wave function:
$\psi(x,t)=\rho(x,t)e^{i \omega(x,t)}$.

In the previous section we have been able to give the conditions
for the equation to have the P-property. Now we  ask, under these
conditions,  whether we can transform \eqref{VCS} to the standard
NLS equation which is an integrable one from a group-theoretical
point of view. This is equivalent to the condition that the
equation under study has a five-dimensional symmetry algebra which
is isomorphic to that of the standard NLS equation. Equivalence
(or allowed) transformations will be used to produce
 the transformations mapping \eqref{VCS} to the NLS equation.

In Section 1, from the Painlev\'e test it turned out that $f$ and $g$ have to be  real functions. When $f$ is a real function, \eqref{qdenk}
is equivalent to
\begin{equation}\label{r} R^2(x,t)=\frac{R_0(t)}{X_x},
\qquad\theta_x=-\frac{1}{2f}\frac{X_t}{X_x},
\end{equation}
where $R_0(t)$ arbitrary. For the equation \eqref{VCS} to have a
five-dimensional symmetry algebra,  we only need to  set
\begin{subequations}\label{fgh}
\begin{equation}\label{f}
\tilde{f}(\tilde{x},\tilde{t})=f(x,t)\frac{X_x^2}{\dot{T}}=1,
\end{equation}
\begin{equation}\label{g}
\tilde{g}(\tilde{x},\tilde{t})=g(x,t)\frac{R^2(x,t)}{\dot{T}}=\epsilon,
\quad \epsilon=\pm1,
\end{equation}
\begin{equation}\label{hh}
\tilde{h}(\tilde{x},\tilde{t})=0.
\end{equation}
\end{subequations}
Notice that we have not included the constant
${i}\tilde{g_2}$ in the righthand side of \eqref{g} since
the functions occurring on the left are real. These equations are
going to  provide us the transformation functions $R,\theta, X$
and $T$.

We recall that we have succeeded  in writing the potential $h$ in
terms of $f$ and $g$ for two (not distinct) cases.

\paragraph{(i)} The case when $f(x,t)g^2(x,t)=K(t)$.

In the following calculations we used the substitution
$f\rightarrow K/g^2$. First solving \eqref{f} we find that
\begin{equation}
X_x=\epsilon_1\sqrt{\frac{\dot{T}}{K}}g, \qquad X(x,t)=
\epsilon_1\sqrt{\frac{\dot{T}}{K}}\int g dx+\xi(t), \quad
\epsilon_1=\pm1
\end{equation}
with an arbitrary $\xi$. From the second equality \eqref{g} we
easily find
\begin{equation}\label{Re}
R(x,t)=\left(\epsilon\frac{ \dot{T}}{g}\right)^{1/2}.
\end{equation}
Next we make use of the equations in \eqref{r}. The first equation
means that the product $R^2(x,t)X_x(x,t)$ should be a function of
$t$, which is indeed the case. Integration of the second equation
determines $\theta$:
\begin{equation}\label{teta}
\theta(x,t)=-\frac{1}{8K}\left(\ln\frac{\dot{T}}{K}\right)_t\;(\int
g dx)^{2}-\frac{1}{2K}\int g\,(\int g_t dx)
dx-\frac{\epsilon_1\dot{\xi}}{2\sqrt{\dot{T}K}}\int g dx+\eta(t),
\end{equation}
where $\eta$ is an arbitrary function. The remaining equation to be solved is
\eqref{hh}. We are going to make use of
\eqref{h2},\eqref{h1},\eqref{Re},\eqref{teta} in the formula
\eqref{h}. From the real and imaginary part of \eqref{hh} we have,
respectively,
\begin{eqnarray}\label{real}
\begin{split}
&\left(\delta_1-\dot{\eta}-\frac{\dot{\xi}^2}{4\dot{T}}\right)+\left(\delta_0+\frac{\epsilon_1\ddot{\xi}}{2\sqrt{K\dot{T}}}-
\frac{\epsilon_1\dot{\xi}\ddot{T}}{2\sqrt{K}\dot{T}^{3/2}}\right)\int g dx\\
&+\frac{1}{8K}\left(\frac{\dddot{T}}{\dot{T}}-\frac{3}{2}\left(\frac{\ddot{T}}{\dot{T}}\right)^2+
\frac{\ddot{K}}{K}-\frac{1}{2}\left(\frac{\dot{K}}{K}\right)^2+4\gamma\frac{\dot{K}}{K}+8\gamma^2+4\dot{\gamma}\right)\left(\int
g dx\right)^2=0,
\end{split}
\end{eqnarray}
\begin{equation}
4\gamma(t)+\frac{\dot{K}}{K}+\frac{\ddot{T}}{\dot{T}}=0.
\end{equation}
The second equation is easy to integrate and at once yields the
time transformation
\begin{equation}\label{Te}
T(t)=T_1\int\frac{e^{-4\int\gamma dt}}{K(t)}dt+T_2, \qquad
T_1,T_2 \; \text{constants}.
\end{equation}
Once $f,g$ and $h$ are given by a specific case of \eqref{VCS},
one can first check the P-property conditions and after that
determine the function $T$, in principle. With this form
of $T$ the last term in \eqref{real} vanishes, but the first and
the second ones do not. However, we can set them equal to zero by choosing
$\xi$ and $\eta$ as solutions of the equations
\begin{subequations}\label{ksieta}
\begin{equation}\label{ksi}
\delta_0(t)+\frac{\epsilon_1\ddot{\xi}}{2\sqrt{K\dot{T}}}-
\frac{\epsilon_1\dot{\xi}\ddot{T}}{2\sqrt{K}\dot{T}^{3/2}}=0,
\end{equation}
\begin{equation}\label{eta}
\delta_1(t)-\dot{\eta}-\frac{\dot{\xi}^2}{4\dot{T}}=0.
\end{equation}
\end{subequations}
Having found $T$ we can arrange \eqref{ksi} as
\begin{equation}
\delta_0(t)+\frac{\epsilon_1}{2}\sqrt{\frac{\dot{T}}{K}}\left(\frac{\dot{\xi}}{\dot{T}}\right)_t=0,
\end{equation}
of which integration twice yields
\begin{equation}
\xi(t)=-2\epsilon_1\int\dot{T}(\int
\sqrt{\frac{K}{\dot{T}}}\,\delta_0(t) dt)dt+\xi_0T(t)+\xi_1,
\end{equation}
where $\xi_0,\xi_1$ are constants. From \eqref{eta} follows
directly
\begin{equation}
\eta(t)=\int\delta_1(t)dt-\int
\dot{T}(\int\sqrt{\frac{K}{\dot{T}}}\delta_0dt)^2dt+\epsilon_1\xi_0\int
\dot{T}(\int\sqrt{\frac{K}{\dot{T}}}\delta_0dt)dt-\frac{\xi_0^2}{4}T+\theta_0,
\end{equation}
with $\theta_0$ constant. So far we have all the information to write the general
transformation formula. $T$ and $R$ are given by \eqref{Te} and
\eqref{Re}, respectively. For the space variable transformation we have
\begin{equation}
X(x,t)=\epsilon_1\sqrt{\frac{\dot{T}}{K}}\int g
dx-2\epsilon_1\int\dot{T}(\int \sqrt{\frac{K}{\dot{T}}}\delta_0(t)
dt) dt+\xi_0T(t)+\xi_1, \quad \epsilon_1=\pm1.
\end{equation}
The phase of $Q$ is obtained in the form
\begin{eqnarray}
\begin{split}
\theta(x,t)=&-\frac{1}{8K}\left(\ln\frac{\dot{T}}{K}\right)_t\;(\int
g dx)^{2}-\frac{1}{2K}\int g\,(\int g_t dx)
dx\\&+\sqrt{\frac{\dot{T}}{K}}(\int\sqrt{\frac{K}{\dot{T}}}\delta_0dt-\frac{\epsilon_1\xi_0}{2})\int
g dx+\int\delta_1dt\\&-\int
\dot{T}(\int\sqrt{\frac{K}{\dot{T}}}\delta_0dt)^2dt+\epsilon_1\xi_0\int
\dot{T}(\int\sqrt{\frac{K}{\dot{T}}}\delta_0dt)dt-\frac{\xi_0^2}{4}T+\theta_0.
\end{split}
\end{eqnarray}

Now we make use of the formula we have found for $T$ and give the
transformation functions together with \eqref{Te} in its most
explicit form. With $\Gamma(t)=2\int \gamma(t)dt$ we have
\begin{align}\label{sonX}
\nonumber X(x,t)=\,&\epsilon_1\sqrt{T_1}\frac{e^{-\gt}}{K}\int g
dx-2\epsilon_1\sqrt{T_1}\int \frac{e^{-2\gt}}{K} (\int
e^{\gt}K\delta_0 dt)dt\\
   &+\xi_0T_1\int\frac{e^{-2\gt}}{K}dt+\xi_1,\\  \label{sonR}
R(x,t)=\,&e^{-\gt} \left(\frac{\epsilon T_1}{gK}\right)^{1/2},\\
\label{sonTT} \nonumber
\theta(x,t)=\,&\frac{1}{4K}(2\gamma+\frac{\dot{K}}{K})(\int g
dx)^{2}-\frac{1}{2K}\int g\,(\int g_t dx) dx-\int
\frac{e^{-2\gt}}{K} (\int e^{\gt}K\delta_0
dt)^2dt\\
\nonumber &+\frac{e^{-\gt}}{K}(\int e^{\gt}K\delta_0
dt-\frac{\epsilon_1\xi_0\sqrt{T_1}}{2})\int g dx
+ \int\delta_1dt-\frac{\xi_0^2T_1}{4}\int \frac{e^{-2\gt}}{K} dt\\
&+\epsilon_1\xi_0\sqrt{T_1} \int \frac{e^{-2\gt}}{K} (\int
e^{\gt}K\delta_0 dt)dt+\theta_0.
\end{align}

\paragraph{(ii)} The case when  $f(x,t)=f(t),\, g(x,t)=g(t)$.

Although this case is already included in (i) we intend to present the
results since a wide literature is devoted to the study of equations
with this special form of the coefficients. In this case, eqs.
\eqref{ksieta} are not imposed but rather satisfied identically. Transformations can be directly obtained
from \eqref{Te},\eqref{sonX}-\eqref{sonTT} via the substitution
\begin{equation}
K(t)=f(t)g^2(t), \quad \gamma(t)=\beta(t)-\frac{\dot{g}(t)}{g(t)},
\quad \delta_0(t)=\frac{H_1(t)}{g(t)}, \quad \delta_1(t)=H_2(t).
\end{equation}
Allowed transformations are found to be, with $B(t)=2\int
\beta(t)dt$,
\begin{align}\label{sonTTT}
T(t)=&T_1\int\frac{g^2}{f}e^{-2B}dt+T_2, \\ \label{sonXX}
\nonumber X(x,t)=&\epsilon_1\sqrt{T_1}\frac{g}{f} e^{-B}
x-2\epsilon_1\sqrt{T_1}\int \frac{g^2}{f}
e^{-2B} (\int\frac{f}{g} e^{B}H_1 dt)dt\\
     &+\xi_0T_1\int \frac{g^2}{f}e^{-2B} dt+\xi_1,\\  \label{sonRR}
R(x,t)=&e^{-B} \left(\epsilon T_1\frac{g}{f}\right)^{1/2},\\
\label{sonTTTT} \nonumber
\theta(x,t)=&\frac{1}{4f}(2\beta+(\ln\frac{f}{g})_t)x^2+\frac{g}{f}e^{-B}(\int
\frac{f}{g} e^{B}H_1 dt-\frac{\epsilon_1\xi_0\sqrt{T_1}}{2})x
\\
\nonumber &-\int \frac{g^2}{f} e^{-2B} (\int \frac{f}{g} e^{B}
H_1 dt)^2dt
-\frac{\xi_0^2T_1}{4}\int \frac{g^2}{f} e^{-2B} dt\\
        &+\epsilon_1\xi_0\sqrt{T_1}
\int \frac{g^2}{f} e^{-2B} (\int \frac{f}{g} e^{B}H_1 dt) dt+
\int H_2dt+\theta_0.
\end{align}

\begin{rmk}
 For $f=f_1+i f_2$, one can set
$X_x^2={\dot{T}}/{f_1}$ and this normalizes $f$ through
\eqref{fg} into $\tilde{f}(x,t)=1+i \tilde{f}_2(x,t)$
\cite{Winternitz93}. Thus, without losing any generality,  we could
equivalently investigate the equation
\begin{equation}
i\psi_t+(1+i f_2)\psi_{xx}+g(x,t) |\psi|^2 \psi+h(x,t) \psi=0.
\end{equation}
From Painlev\'e analysis of Section 2  would immediately follow
$f_2=0$ and  $g=g(t)$ as a real-valued function  and the potential
components
\begin{equation}\label{pot}
h_{1}(x,t)=(\beta^2+\frac{\dot{\beta}}{2}-\beta\frac{\dot{g}}{g}+\frac{g}{4}(\frac{1}{g})_{tt})x^2+H_1(t)x+H_2(t), \quad h_2(x,t)=\beta(t).
\end{equation}
Transformation formulae would considerably simplify in this
setting. Yet, our results will remain  useful when one begins with
an equation with $f\neq1$.
\end{rmk}

\section{Integrability of a Gross-Pitaevskii Equation}
We apply our results obtained in previous sections to the equation
\begin{equation}\label{GPGP}
i\psi_t+\psi_{xx}+g(x,t) |\psi|^2 \psi+kx^2 \psi=0,
\end{equation}
where $g\in\mathbb{C}$ and $k\in\mathbb{R}$. This equation governs
the dynamics of a Bose-Einstein condensate and its integrability
was studied in \cite{ZLC08}  through the Painlev\'e analysis. The
equation was shown to pass the Painlev\'e test for PDEs (WTC test)
when the coefficient $g$ satisfies
\begin{equation}\label{difg}
g\,\ddot{g}-2\dot{g}^2+4k g^2=0
\end{equation}
and has the special form
\begin{equation}\label{sxt}
g(x,t)=g(t)=\frac{2g_0e^{\pm2\sqrt{k}t}}{Ae^{\pm4\sqrt{k}t}-B},
\end{equation}
where $A,B,g_0$  are arbitrary constants. Subsequently, by a proper transformation the equation is
transformed to standard integrable nonlinear Schr\"{o}dinger equation.

We can recover the equation \eqref{difg} using our results of Section 1. Then, without considering results of
the Painlev\'e analysis, we will try to transform the equation into the
standard NLS equation based on its group-theoretical properties. To this end, we shall require that the equation under study has a five-dimensional symmetry algebra. We then show
that this invariance requirement  will force $g$ to have exactly the same
form obtained by the Painlev\'e approach.

Identifying \eqref{GPGP} with \eqref{VCS}, we see that for the
Gross-Pitaevskii equation $f_1=1, \; f_2=0 $, coefficient of the
cubic term is the same complex function  $g(x,t)$ and
$h_1(x,t)=kx^2, \; h_2=0$. Since $f$ is a constant (which is equal
to $1$), $g$ is necessarily a function of $t$ only. We make use of
\eqref{ref} and obtain the same equation \eqref{difg} for $g$,
under which equation \eqref{GPGP} does have the P-property. If $g$ is
taken as a solution to \eqref{difg}, the GP equation transforms to
standard NLS equation through the transformation formulae
\eqref{sonTTT}-\eqref{sonTTTT}.

Existence of a transformation into standard NLS equation naturally motivates
the question whether \eqref{GPGP} can have a five-dimensional
symmetry algebra for some form of coefficient function $g(x,t)$,
in particular the one given in \eqref{sxt}. We try to transform
\eqref{GPGP} directly to the NLS equation via \eqref{r} and \eqref{fgh}.
From \eqref{f} we have
\begin{equation}
X(x,t)=\epsilon_1\sqrt{\dot{T}}x+\xi(t)
\end{equation}
and \eqref{r} implies $R(x,t)=R(t)$ and
\begin{equation}
\theta(x,t)=-\frac{1}{8}\frac{\ddot{T}}{\dot{T}}\,x^2-\frac{\epsilon_1\dot{\xi}(t)}{2\sqrt{\dot{T}}}\,x+\eta(t).
\end{equation}
\eqref{hh} has real and imaginary parts, both to be set to zero. From the imaginary part we have $\frac{\dot{R}}{R}+\theta_{xx}=0$, which integrates to give
\begin{equation}\label{R}
R(t)=R_0[\dot{T}(t)]^{1/4}, \qquad  R_0=\mathrm{constant}.
\end{equation}
From the real part we have $\theta_x^2+\theta_t-kx^2=0$ and this gives  differential conditions on  $T(t),\xi(t)$ and $\eta(t)$:
\begin{subequations}
\begin{equation}\label{T}
\frac{\dddot{T}}{\dot{T}}-\frac{3}{2}\left(\frac{\ddot{T}}{\dot{T}}\right)^2+8k=0,
\end{equation}
\begin{equation}
\ddot{\xi}\dot{T}-\dot{\xi}\ddot{T}=0,
\end{equation}
\begin{equation}
\dot{\xi}^2+4\dot{\eta}\dot{T}=0.
\end{equation}
\end{subequations}
The second equation yields
\begin{equation}
\xi(t)=\xi_0 T(t)+\xi_1
\end{equation}
with $\xi_0,\xi_1$ constants, while from the third it follows that  $$\eta(t)=-\frac{1}{4}\xi_0^2\;T(t)+\theta_0$$ with the constant $\theta_0$. These determine $\theta$ in terms of $T$:
\begin{equation}
\theta(x,t)=-\frac{1}{8}\frac{\ddot{T}(t)}{\dot{T}(t)}x^2-\frac{\epsilon_1\xi_0}{2}\dot{T}^{1/2}(t)\,x-\frac{\xi_0^2}{4}T(t)+\theta_0
\end{equation}
There remains the solution of \eqref{T}, which is recognizable to be a Schwarzian differential equation which we rewrite as
\begin{equation}\label{sch}
\left\{T,t\right\}+8k=0,
\end{equation}
where $\{T,t\}$ denotes the Schwarzian derivative of $T$ with respect to the variable $t$. By the Schwarz theorem it is well-known that its solution  is given by a ratio $T(t)=\frac{\sigma_1(t)}{\sigma_2(t)}$, where $\sigma_1$ and $\sigma_2$ are linearly independent solutions of
\begin{equation}\label{sigma}
\ddot{\sigma}(t)-4k\sigma(t)=0.
\end{equation}
For a derivation of this result, the reader is referred to
\cite{Hille}. An alternative formulation based on a unimodular
group invariance can be found in \cite{Olver95}. Furthermore, if
the constants $T_1,...,T_4$ satisfy $T_2T_3-T_1T_4\neq0$,
$T_1\sigma_1(t)+T_2\sigma_2(t)$ and
$T_3\sigma_1(t)+T_4\sigma_2(t)$ are also linearly independent.
Hence
\begin{equation}
T(t)=\frac{T_1\sigma_1(t)+T_2\sigma_2(t)}{T_3\sigma_1(t)+T_4\sigma_2(t)}
\end{equation}
is a general solution to \eqref{sch}. Linearly independent
solutions of \eqref{sigma} have constant Wronskian, so, without
loss of generality, we can arrange to have
\begin{equation}\label{wr}
W[T_1\sigma_1(t)+T_2\sigma_2(t),T_3\sigma_1(t)+T_4\sigma_2(t)]=(T_1T_4-T_2T_3)W[\sigma_1(t),\sigma_2(t)]=-1,
\end{equation}
which provides $T$ as a solution with three arbitrary constants to
the third-order equation \eqref{sch}. After further calculations,
we are going to see that this choice of the Wronskian will enable
us to obtain the function $g(x,t)$ in its  simplest form.

Let $k>0$. From \eqref{sigma} we have
$\sigma_1(t)=\cosh2\sqrt{k}t$ and  $\sigma_2(t)=\sinh2\sqrt{k}t$
and \eqref{wr} imposes $T_2T_3-T_1T_4=\frac{1}{2\sqrt{k}}$. We
write
\begin{equation}
T(t)=\frac{T_1\cosh2\sqrt{k}t+T_2\sinh2\sqrt{k}t}{T_3\cosh2\sqrt{k}t+T_4\sinh2\sqrt{k}t}
\end{equation}
and
\begin{equation}
\dot{T}(t)=[\,T_3\cosh2\sqrt{k}t+T_4\sinh2\sqrt{k}t]^{-2}.
\end{equation}

For $k<0$, we choose solutions of \eqref{sigma} as
$\sigma_1(t)=\cos2\sqrt{-k}t$ and $\sigma_2(t)=\sin2\sqrt{-k}t$
and write
\begin{equation}
T(t)=\frac{T_1\cos2\sqrt{-k}t+T_2\sin2\sqrt{-k}t}{T_3\cos2\sqrt{-k}t+T_4\sin2\sqrt{-k}t}
\end{equation}
and
\begin{equation}
\dot{T}(t)=[\,T_3\cos2\sqrt{-k}t+T_4\sin2\sqrt{-k}t]^{-2},
\end{equation}
with the similar condition  $T_2T_3-T_1T_4=\frac{1}{2\sqrt{-k}}$.

Now we have determined all the unknown functions of the
equivalence transformation. Our aim was to determine $g(x,t)$ so
that \eqref{GPGP} has a five-dimensional symmetry algebra. From \eqref{R} and \eqref{g} we find
\begin{equation}\label{tg}
g(x,t)=g(t)=\frac{\epsilon}{R_0^2}\;\dot{T}^{1/2}.
\end{equation}
The following summarizes possible forms of $g(x,t)$:
\begin{equation}\label{hyp}
g_1(t)=\frac{\epsilon}{R_0^2}\,[\,T_3\cosh2\sqrt{k}t+T_4\sinh2\sqrt{k}t]^{-1},
\qquad k>0,
\end{equation}
\begin{equation}\label{trg}
g_2(t)=\frac{\epsilon}{R_0^2}\,[\,T_3\cos2\sqrt{-k}t+T_4\sin2\sqrt{-k}t]^{-1},
\qquad k<0.
\end{equation}
Now we would like to compare the results obtained by symmetry
consideration  with those given in \cite{ZLC08}. In a
search for the function $g(x,t)$ through Painlev\'e analysis, authors
of \cite{ZLC08} obtained the differential equation \eqref{difg}
and write \eqref{sxt} as its solution. Setting
\begin{equation}
\dot{T}=L g^2,\quad L=\text{constant}
\end{equation}
taking the relation \eqref{tg} between $T$ and $g$ into account,
and plugging it in \eqref{T}, we obtain exactly the equation
\eqref{difg} for $g$. This implies that \eqref{sxt} ought to be
contained in $g(x,t)$ obtained by our approach. Indeed, if we
choose
$$T_3=\frac{R_0^2}{\epsilon}\frac{A-B}{2g_0}\quad
\text{and} \quad T_4=\pm\frac{R_0^2}{\epsilon}\frac{A+B}{2g_0}$$
in \eqref{hyp}, we see that  it transforms to \eqref{sxt}. We note
that the solution \eqref{sxt} does not include  the case when
$k<0$. The possibility for $g$ given by \eqref{trg} should be
added as an integrable case.

In fact, \eqref{T} is nothing but a special case of the third
term which  vanished identically in \eqref{real} because
at the beginning of Section 3 we  assumed that the equation have
the P-property; i.e., \eqref{h1} holds. But now, since we have not
taken  the P-property into consideration, it is obtained as an equation that has
to be satisfied. We show that it is equivalent to the
condition that the equation has the P-property.

To illustrate our results in a tidy form we introduce the
functions
\begin{equation}
U(t)=\begin{cases}
   T_1\cosh2\sqrt{k}t+T_2\sinh2\sqrt{k}t,   & \text{for $k >0$} \\
  T_1\cos2\sqrt{-k}t+T_2\sin2\sqrt{-k}t, & \text{for $k<0$,} \\
      \end{cases}
\end{equation}
\begin{equation}
V(t)=\begin{cases}
   T_3\cosh2\sqrt{k}t+T_4\sinh2\sqrt{k}t,   & \text{for $k >0$} \\
  T_3\cos2\sqrt{-k}t+T_t\sin2\sqrt{-k}t, & \text{for $k<0$}. \\
      \end{cases}
\end{equation}
In conclusion, we have shown that, if $g(x,t)$ of \eqref{GPGP} is one of $g_j(t),
\; j=1,2$, using the transformation
\begin{eqnarray}
\begin{split}
\tilde{t}(t)&=T(t)=\frac{U}{V},\\
\tilde{x}(x,t)&=X(x,t)=\frac{\epsilon_1}{V}\,x+\xi_0\frac{U}{V}+\xi_1,\\
R(x,t)&=\frac{R_0}{V^{1/2}},\\
\theta(x,t)&=\frac{\dot{V}}{4V}\,x^2-\frac{\epsilon_1\xi_0}{V}\,x-\frac{\xi_0^2}{4}\frac{U}{V}+\theta_0,
\end{split}
\end{eqnarray}
we can convert \eqref{GPGP} to the equation
\begin{equation}\label{NLS}
i\tilde{\psi}_{\tilde{t}}+\tilde{\psi}_{\tilde{x}\tilde{x}}+\epsilon|\tilde{\psi}|^2\tilde{\psi}=0.
\end{equation}
Thus, if $\tilde{\psi}(\tilde{x},\tilde{t})$ is any solution of
\eqref{NLS}, then for $g_j(t)$, $j=1,2$
\begin{equation}
\psi(x,t)=R(x,t)e^{i \theta(x,t)}\;\tilde{\psi}\left(X(x,t),T(t)\right)
\end{equation}
is a solution of \eqref{GPGP}.

As a by-product,  we can transform the symmetry algebra
\eqref{alg} by the equivalence transformations to obtain a basis
for the symmetry algebra of \eqref{GPGP} as follows:
\begin{eqnarray}\nonumber
X_1&=&V^2\dt+(-\xi_0V+V\dot{V}x)\dx-\frac{1}{2}\rho
V\dot{V}\dr+\left(\frac{\xi_0^2}{4}-\frac{\xi_0}{2}\dot{V}x+\frac{1}{4}(\dot{V}^2+V\ddot{V})x^2\right)\dw,\\\nonumber
X_2&=&V\dx+\frac{1}{2}(-\xi_0+\dot{V}x)\dw,\\
X_3&=&\dw, \\\nonumber
X_4&=&U\dx+\frac{1}{2}(\xi_1+\frac{1+U\dot{V}}{V}x)\dw,\\\nonumber
X_5&=&UV\dt+\left(-\frac{\xi_0}{2}U+\frac{\xi_1}{2}V+(\frac{1}{2}+U\dot{V})x\right)\dx-\frac{1}{2}\rho(1+U\dot{V})\dr\\\nonumber
&+&\left(-\frac{\xi_0\xi_1}{4}+(-\frac{\xi_0}{4}\frac{1}{V}+\frac{\xi_1}{4}\dot{V}-\frac{\xi_0}{4}\frac{U\dot{V}}
{V})x+\frac{1}{4}(\frac{\dot{V}}{V}+\frac{U\dot{V}^2}{V}+U\ddot{V})x^2\right)\dw.
\end{eqnarray}

\section{Summary}

We took the approach of Painlev\'e test combined with the symmetry properties as a preliminary attempt  towards integrability of the most general  variable coefficient NLS equation. The other  tool which plays a central role  in the analysis is the notion of equivalence transformations relating the VCNLS equation to its canonical form. Through the Painlev\'e test we obtained restrictions on coefficients of the equation under study. This motivated us to  construct suitable point transformations taking our original equation to  standard cubic NLS equation.
As confirmed by other similar studies existing in literature, equations passing Painlev\'e test possess a Lax pair which is usually considered as an essential characteristic of complete integrability.  We see that this is also the case for the coefficients singled out by the Painlev\'e test.

The results of this paper can be used to obtain exact (soliton) solutions of the VCNLS equation, when the coefficients have some specific form, from those of the integrable NLS equation by point transformations whose precise forms are presented throughout the paper.


\begin{thebibliography}{99}

\bibitem{Gungor96}F. G\"{u}ng\"{o}r, M. Sanielevici and P. Winternitz, 1996, On the integrability properties of variable coefficient Korteweg-de Vries equations, \emph{Can. J. Phys.}, 74, 676-684.

\bibitem{Gungor01-3}
F.~G{\"u}ng{\"o}r and P.~Winternitz, 2002, Generalized
Kadomtsev-petviashvili equation with an
  infinite dimensional symmetry algebra.
\emph{J. Math. Anal. Appl.},  276, 314-328.


\bibitem{Zhu09}          Zhu, J-M., Liu, Y-L., 2009, Some exact solutions of variable coefficient cubic-quintic nonlinear Schr\"{o}dinger
                                equation with an external potential,
                           \emph{Commun. Theor. Phys.}, 51, (3), 391-394.
\bibitem{Hao04}  Hao, R., Lu, L., Zhongao, L. et al., 2004. A new approach to exact soliton solutions and soliton interaction for the
                              nonlinear Schr\"{o}dinger equation with variable coefficients,
                           \emph{Optics Communications}, 236, 79-86.


\bibitem{Zhang07}  Zhang, J-L., Li, B-A., Wang, M-L., 2009, The exact solutions and the relevant constraint conditions for two nonlinear Schr\"{o}dinger equations with variable coefficients,     \emph{Chaos, Solitons and Fractals}, 39, 858-865.

\bibitem{Lu07}  L\"{u}, X., Zhu, H-W. et al., 2007, Soliton solutions and a B\"{a}cklund transformation
                                for a generalized nonlinear Schr\"{o}dinger equation  with variable coefficients from optical fiber communications,
                           \emph{J. Math. Anal. Appl.}, 336, 1305-1315.

\bibitem{Lu08}  L\"{u}, X., Zhu, H-W. et al., 2008, Multisoliton solutions in terms of double
                                    Wronskian determinant for a generalized
                                        variable-coefficient nonlinear Schr\"{o}dinger
                                        equation from plasma physics, arterial
                                            mechanics, fluid dynamics and optical
                                                    communications,
                           \emph{Annals of Physics}, 323, 1947-1955.

\bibitem{Kruglov03}  Kruglov, V.I., Peacock, A.C., Harvey, C.D., 2003, Exact self-similar solutions of the generalized nonlinear Schr\"{o}dinger equation
                                    with distributed coefficients,
                           \emph{Pyhs. Rev. Let.}, 90, (11), 113902.


\bibitem{Liu07}  Liu, X-Q., Yan, Z-L., 2007, Some exact solutions of the variable coefficient
                                    Schr\"{o}dinger equation,
                           \emph{Comm. in N. Sci. and Num. Sim.}, 12, 1355-1359.

\bibitem{Yan07}   Yan, Z., Liu, X., Wang, L., 2007, The direct symmetry method and its application
                                         in variable coefficients Schr\"{o}dinger equation,
                           \emph{Applied Mathematics and Computation}, 187, 701-707.


 \bibitem{Liu04}  Liu, X.Q. et al., 2004, Soliton solutions in linear magnetic field and
                                        time-dependent laser field,
                           \emph{Comm. in N. Sci. and Num. Sim.}, 9, 361-365.

 \bibitem{Beitia08}  Belmonte-Beitia, J., Perez-Garc\'{\i}a, V.M. et al., 2008, Localized nonlinear waves in systems with time- and space-modualated
                                nonlinearities,
                           \emph{Phys. Rev. Let.}, 100, 164102.

 \bibitem{Serkin07}  Serkin, V.N., Hasegawa, A., Belyaeva, T.L., 2007, Nonautonomous solitons in external potentials,
                           \emph{Phys. Rev. Let.}, 98, 074102.

\bibitem{Gurses08}  G\"{u}rses, M., 2008, Integrable nonautonomous nonlinear Schr\"{o}dinger equations,
                           \emph{arXiv e-print}, 0704.2435v2[nlin.SI].

\bibitem{Kundu09}  Kundu, A. 2009, Integrable nonautonomous nonlinear Schrodinger equations are equivalent to the standard autonomous equation,
                           \emph{Phys. Rev.  E}, 79, 015601(R); arXiv:0809.1924 [nlin.SI]
.

\bibitem{Steeb84}  Steeb, W-H., Kloke, M.,Spieker, B.M., 1984, Nonlinear Schr\"{o}dinger equation, Painlev\'e test, B\"{a}cklund transformation and solutions,
                           \emph{J. Phys. A:Math. Gen.}, 17, L825-L829.

\bibitem{ZLC08}  Zhao, D., Luo, H-G., Chai, H-Y., 2008, Integrability of the  Gross-Pitaevskii equation with Feshbach resonance management,
                           \emph{Phys. Let. A}, 372, 5644-5650.

\bibitem{ZHL08}  Zhao, D., He, X-G., Luo, H-G., 2008, From canonical to nonautonomous solitons,
                           \emph{arXiv e-print}, 0807.1192v1 [nlin.PS].

\bibitem{LZHL08}  Luo, H-G., Zhao, D. et al., 2008, Dissipative solutions stabilized by a quantum Zeno-like effect,
                           \emph{arXiv e-print}, 0808.3437v2 [nlin.PS].



\bibitem{Winternitz93}  Gagnon, L., Winternitz, P., 1993, Symmetry classes of variable coefficient Schr\"{o}dinger equations,
                           \emph{J. Phys. A:Math.Gen.}, (26) 7061-7076.
\bibitem{Brugarino10}  Brugarino, T., Sciacca, M.,  2010, Integrability of a nonlinear Schr\"{o}dinger equation,   \emph{Journal of Math. Phys.}, (51), 093503.

\bibitem{Li07}  Li, J. et al.,  2007, Soliton-like solutions of a generalized variable coefficient higher order nonlinear Sch\"{o}dinger equation
                                       from inhomogeneous optical fibers with symbolic computation,   \emph{Journal of Phys. A: Math. and Theor.}, (40), 13299-13309.


\bibitem{Zhao09}  He, X-G et al., 2009, Engineering
                              integrable nonautonomous nonlinear Schr\"{o}dinger equations,
                           \emph{Physical Review E}, (79), 056610.


\bibitem{Clarkson93}  Ablowitz, M.J., Clarkson, P.A., 1999, Solitons, nonlinear evolution equations and inverse scattering,
                           \emph{Cambridge Univ. Press}.




\bibitem{Hille}
E. Hille.
\newblock {\em Ordinary Differential Equations in the Complex Plane}.
\newblock Dover Publications,  1997.



\bibitem{Olver95}
P.J. Olver.
\newblock {\em Equivalence, Invariants and Symmetry}.
\newblock Cambridge University Press, Cambridge, 1995.





\end{thebibliography}

\end{document}